\documentclass[proceedings]{stacs}
\stacsheading{2009}{123--134}{Freiburg}
\firstpageno{123}

\usepackage{amsmath,amssymb}
\usepackage[OT1]{fontenc}
\usepackage[latin1]{inputenc}
\usepackage{comment}
\usepackage[ruled,lined]{algorithm2e}
\usepackage{multicol}
\usepackage{subfigure}

\def \cg {[ \negthinspace [ }
\def \cd {] \negthinspace ] }
\def \quotient#1 {/\negthinspace{#1} }
\def\cqfd{\skip10=\parfillskip\parfillskip=0pt
\enspace\hfill\symbolecqfd\par\parfillskip=\skip10\par\medskip}
\def\symbolecqfd{\rlap{$\sqcap$}$\sqcup$}

\def\iver#1{\cg {#1}\cd }

\newenvironment{mainproof}{\rm \trivlist \item[\hskip \labelsep{\bf
     Proof of the main theorem: }]}{\cqfd\endtrivlist}

\def \p {\cdot}

\def \A {\mathcal{A}}

\def \F {\mathcal{F}}
\def \G {\mathcal{G}}

\def \T {\mathcal{T}}

\def \N {\mathbb{N}}
\def \M {\textsc{Moore}}
\def \O {\mathcal{O}}

\title{On the average complexity of Moore's state minimization algorithm}
\author[ref1]{F. Bassino}{Fr\'ed\'erique Bassino}
\author[ref2]{J. David}{Julien David}
\author[ref2]{C. Nicaud}{Cyril Nicaud}
\address[ref1]{LIPN UMR 7030, Universit\'e Paris 13 - CNRS, 99, avenue
  Jean-Baptiste Cl\'ement, 93430 Villetaneuse, France.}
\email{Frederique.Bassino@lipn.univ-paris13.fr}
\address[ref2]{Institut Gaspard Monge, Universit\'e Paris Est, 77454
  Marne-la-Vall\'ee Cedex 2, France}
\email{Julien.David@univ-paris-est.fr, Cyril.Nicaud@univ-paris-est.fr}
\keywords{finite automata, state minimization, Moore's algorithm, average
complexity} 
\subjclass{F.2 Analysis of algorithms and problem complexity}

\begin{document}
\begin{abstract} 
  We prove that, for any arbitrary finite alphabet and for the uniform
  distribution over deterministic and accessible automata with $n$
  states, the average complexity of Moore's state minimization
  algorithm is in $\O(n \log n)$. Moreover this bound is tight in the
  case of unary automata.
\end{abstract}

\maketitle

\section{Introduction} 
Deterministic automata are a convenient way to represent regular
languages that can be used to efficiently perform most of usual
computations involving regular languages. Therefore finite state
automata appear in many fields of computer science, such as
linguistics, data compression, bioinformatics, etc. To a given regular
language one can associate a unique smallest deterministic automaton,
called its minimal automaton. This canonical representation of regular
languages is compact and provides an easy way to check equality.
As a consequence, state minimization algorithms that compute the minimal
automaton of a regular language, given by a deterministic automaton, are
fundamental.

Moore proposed a solution \cite{Moore56} that can be seen as a
sequence of partition refinements. Starting from a partition of the
set of states, of size $n$, into two parts, successive refinements
lead to a partition whose elements are the subsets of
indistinguishable sets, that can be merged to form a smaller automaton
recognizing the same language. As there are at most $n-2$ such
refinements, each of them requiring a linear running time, the
worst-case complexity of Moore's state minimization algorithm is
quadratic.  Hopcroft's state minimization algorithm \cite{Hop71} also
uses partition refinements to compute the minimal automaton, selecting
carefully the parts that are split at each step. Using suitable data
structures, its worst-case complexity is in $\O(n\log n)$. It is the
best known minimization algorithm, and therefore it has been
intensively studied, see \cite{bc06,bc04,gri73,knu01} for instance.
Finally Brzozowski's algorithm~\cite{brzo62, cha02} is different from
the other ones. Its inputs may be non-deterministic automata.
It is based on two successive determinization steps, and though its
worst-case complexity is proved to be exponential, it has been noticed
that it is is often sub-exponential in practice. The reader is invited
to consult \cite{Watson01}, which presents a taxonomy of minimization
algorithms, for a more exhaustive list.

\begin{figure}
   \begin{minipage}[c]{.46\linewidth}
   \rotatebox{-90}{ \includegraphics[scale=0.36]{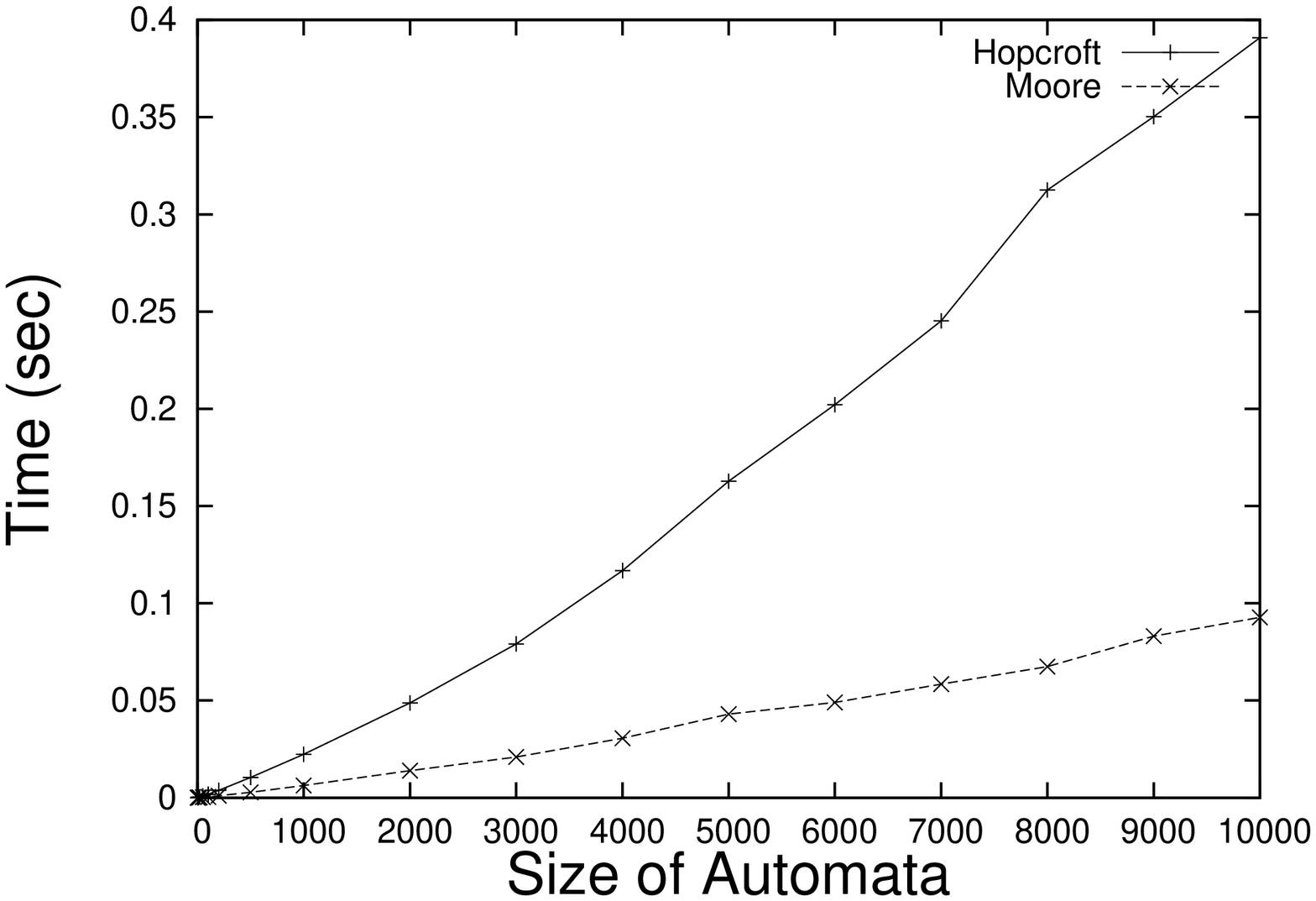}}
     \label{fig:comparison}  
    \caption{Time complexity of  Moore's and Hopcroft's algorithms} 
 \end{minipage} \hfill   
\begin{minipage}[c]{.46\linewidth}
      \rotatebox{-90}{\includegraphics[scale=0.36]{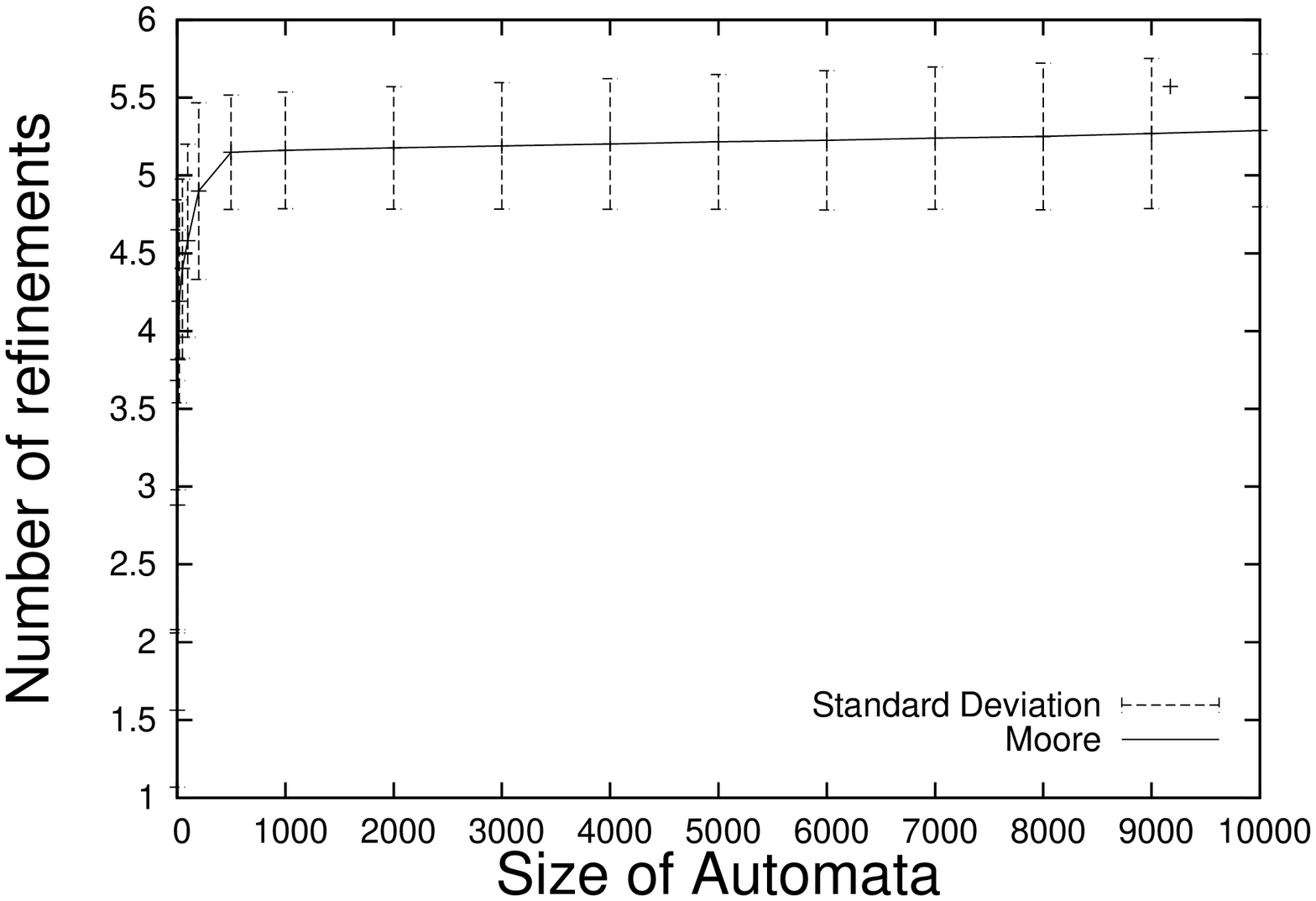}}
    \caption{Number of partition refinements in Moore's algorithm}
\label{fig:iterations} 
\end{minipage}

The results of Fig.\ref{fig:comparison} and Fig.\ref{fig:iterations}
were obtained with the C++ library REGAL (available at: {\tt
http://regal.univ-mlv.fr/}) to randomly generate deterministic
accessible automata \cite{bn07,bdn07,bdn08}.  Each  value
is computed from 20\,000 automata over a $2$-letter alphabet.
\end{figure}

In this paper we study the average time complexity of Moore's
algorithm. From an experimental point of view, the average complexity
of Moore's algorithm seems to be smaller than the complexity of
Hopcroft's algorithm (Fig.\ref{fig:comparison}) and the number of
partition refinements increases very slowly as the size of the input
grows (Fig.\ref{fig:iterations}). In the following we mainly prove
that in average, for the uniform distribution, Moore's algorithm
performs only $\O(\log n)$ refinements, thus its average complexity 
is in $\O(n\log n)$.

After briefly recalling the basics of minimization of automata in
Section~\ref{sec:def}, we prove in Section~\ref{sec:main} that the
average time complexity of Moore's algorithm is $\O(n \log n)$ and
show in Section \ref{sec:unary} that this bound is tight when the
alphabet is unary.  The paper closes with a short discussion about
generalizations of our main theorem to Bernoulli distributions and to
incomplete automata in Section~\ref{sec:discussion}, and the
presentation of a conjecture based on the slow growth of the number
of refinements (Fig.\ref{fig:iterations} when the alphabet is not
unary in Section \ref{sec:conj}.

\section{Preliminaries}\label{sec:def}
This section is devoted to basic notions related to the minimization
of automata. We refer the reader to the literature for more details
about minimization of automata \cite{Hop-Ull, Lothaire3,Woo87}. We
only record a few definitions and results that will be useful for our
purpose.
\subsection{Finite automata}
A {\em finite deterministic automaton} $\A$ is a quintuple
$\A=(A,Q,\p,q_0,F)$ where $Q$ is a finite set of {\em states}, $A$ is
a finite set of {\em letters} called {\em alphabet}, the {\em
  transition function} $\p$ is a mapping from $Q\times A$ to $Q$,
$q_0\in Q$ is the {\em initial state} and $F\subset Q$ is the set of
final states. An automaton is {\em complete} when its transition
function is total.  The transition function can be extended by
morphism to all words of $A^*$: $p\p\varepsilon =p$ for any $p\in Q$
and for any $u,v\in A^*$, $p\p (uv) = (p\p u)\p v$. A word $u\in A^*$
is {\em recognized} by an automaton when $p\p u\in F$.  The set of all words
recognized by $\A$ is denoted by $L(\A)$.  An automaton is {\em
  accessible} when for any state $p\in Q$, there exists a word $u \in
A^*$ such that $q_0\p u = p$.
 
A {\em transition structure} is an automaton where the set of final
states is not specified. Given such a transition structure
$\T=(A,Q,\p,q_0)$ and a subset $F$ of $Q$, we denote by $(\T,F)$ the
automaton $(A,Q,\p,q_0,F)$. For a given deterministic and accessible
transition structure with $n$ states there are exactly $2^n$ distinct
deterministic and accessible automata that can be built from this
transition structure. Each of them corresponds to a choice of set of
final states.

In the following we only consider complete accessible deterministic
automata and complete accessible deterministic transition structures,
except in the presentation of the generalizations of the main theorem
in Section~\ref{sec:discussion}. Consequently these objects will often
just be called respectively {\em automata} or {\em transition
structures}.  The set $Q$ of states of an $n$-state transition
structure will be denoted by $\{1,\cdots,n\}$.

\subsection{Myhill-Nerode equivalence}

Let $\A=(A,Q,\p,q_0,F)$ be an automaton. For any nonnegative integer
$i$, two states $p,q\in Q$ are {\em $i$-equivalent}, denoted by
$p\sim_i q$, when for all words $u$ of length less than or equal to
$i$, $\iver{p\p u\in F} = \iver{q\p u\in F} $, where the Iverson
bracket $\iver{Cond}$ is equal to $1$ if the condition $Cond$ is
satisfied and $0$ otherwise.  Two states are {\em equivalent} when for
all $u\in A^*$, $\iver{p\p u\in F} = \iver{q\p u\in F} $. This
equivalence relation is called {\em Myhill-Nerode equivalence}.  An
equivalence relation $\equiv$ defined on the set of states $Q$ of a
deterministic automaton is said to be {\em right invariant} when
$$\mbox{for all } u\in A^* \mbox{ and all } p,q\in Q, \quad p\equiv
q\Rightarrow p\p u \equiv q\p u.$$ The following proposition
\cite{Hop-Ull, Lothaire3,Woo87} summarizes the properties of
Myhill-Nerode equivalence that will be used in the next sections.

\begin{proposition}\label{nerode}
Let $\A=(A,Q,\p,q_0,F)$ be a deterministic automaton with $n$
states. The following properties hold:
\begin{enumerate}
\item For all $i\in\N$, $\sim_{i+1}$ is a partition refinement of
$\sim_i$, that is, for all $p,q\in Q$, if $p\sim_{i+1} q$ then
$p\sim_i q$.
\item  For all $i\in\N$ and for all $p,q\in Q$, $p\sim_{i+1} q$ if
and only if $p\sim_i q$ and for all $a\in A$, $p\p a \sim_i q\p a$.
\item If for some $i\in\N$ $(i+1)$-equivalence is equal to
  $i$-equivalence then for every $j\geq i$,  $j$-equivalence is
  equal to Myhill-Nerode equivalence.
\item $(n-2)$-equivalence is equal to Myhill-Nerode equivalence.
\item Myhill-Nerode equivalence is right invariant.
\end{enumerate}
\end{proposition}

Let $\A=(A,Q,\p,q_0,F)$ be an automaton and $\equiv$ be a right
invariant equivalence relation on $Q$.  The {\em quotient automaton} of $\A$
by $\equiv$ is the automaton 
\[\left(\A\quotient{\equiv} \right)=(A,Q\quotient{\equiv} ,*,[q_0],\{[f],f\in
F\}),\] where $Q\quotient{\equiv} $ is the set of equivalent classes,
$[q]$ is the class of $q\in Q$, and $*$ is defined for any $a\in A$
and any $q\in Q$ by $[q]*a=[q\p a]$. The correctness of this
definition relies on the right invariance of the equivalence relation
$\equiv$.

\begin{theorem}\label{minimal}
  For any complete, accessible and deterministic automaton $\A$, the
  automaton $\A\quotient{\sim} $ is the unique smallest automaton (in
  terms of the number of states) that recognizes the same language as
  the automaton $\A$. It is called the {\em minimal automaton} of
  $L(\A)$.
\end{theorem}
The uniqueness of the minimal automaton is up to labelling of the
states.  Theorem~\ref{minimal} shows that the minimal automaton is a
fundamental notion in language theory: it is the most space efficient
representation of a regular language by a deterministic automaton, and
the uniqueness defines a bijection between regular language and minimal
automata. For instance, to check whether two regular languages are
equal, one can compare their minimal automata.  It is one of the
motivations for the algorithmic study of the computation of the
minimal automaton of a language.

\subsection{Moore's state minimization algorithm}\label{sec-moore}

In this section we describe the algorithm due to Moore~\cite{Moore56}
which computes the minimal automaton of a regular language represented
by a deterministic automaton. Recall that Moore's algorithm builds
the partition of the set of states corresponding to Myhill-Nerode
equivalence. It mainly relies on properties (2) and (3) of
Proposition~\ref{nerode}: the partition $\pi$ is initialized according
to the $0$-equivalence $\sim_0$, then at each iteration the partition
corresponding to the $(i+1)$-equivalence $\sim_{i+1}$ is computed from
the one corresponding to the $i$-equivalence $\sim_i$ using property
(2). The algorithm halts when no new partition refinement is obtained,
and the result is Myhill-Nerode equivalence according to property
(3). The minimal automaton can then be computed from the resulting
partition since it is the quotient automaton by Myhill-Nerode
equivalence.

\vspace{0.2cm}

\noindent
{\small
  \begin{minipage}{.5\textwidth}%
    \linesnumbered 
    \begin{algorithm}[H]%
      \dontprintsemicolon
      \If{$F=\emptyset$}{
        \Return{$(A,\{1\},*,1,\emptyset)$}\;
      }%
      \If{$F=\{1,\cdots,n\}$}{
        \Return{$(A,\{1\},*,1,\{1\})$}\;
      }%
      \BlankLine
      \ForAll{$p\in \{1,\cdots,n\}$}{
        $\pi'[p] = \iver{p\in F} $\;
      }
      \BlankLine
      $\pi = $ undefined\;
      
      \While{$\pi \neq \pi'$}{
        $\pi = \pi'$\;
        compute the partition $\pi'$ s.t. 
        $\pi'[p]=\pi'[q]$ iff $\pi[p] = \pi[q]$\; 
        and $\forall a\in A$ $\pi[p\cdot a] = \pi[q\cdot a]$\;
      }
      \Return{the quotient of $\A$ by $\pi$}
      \caption{Moore}
    \end{algorithm}
  \end{minipage}%
  \begin{minipage}{.5\textwidth}%
    \linesnumbered 
    \begin{algorithm}[H]%
      \dontprintsemicolon
      \ForAll{$p\in \{1,\cdots,n\}$}{
        $s[p] = (\pi[p],\pi[p\p a_1],\cdots,\pi[p\p a_k])$\;
      }
      \BlankLine
      compute the permutation $\sigma$ that sorts the states according to $s[]$\;
      \BlankLine
       $i=0$\;
      $\pi'[\sigma(1)] = i$\;
      \BlankLine
       \ForAll{$p\in \{2,\cdots,n\}$}{
        \lIf{$s[p]\neq s[p-1]$} { $i=i+1$\;}
        $\pi'[\sigma(p)] = i$\;
      }
      \Return{$\pi'$}
      \caption{Computing $\pi'$ from $\pi$}
    \end{algorithm}%
  \end{minipage}%
}

In the description of Moore's algorithm, $*$ denotes the function such
that $1*a=1$ for all $a\in A$. Lines 1-6 correspond to the special
cases where $F=\emptyset$ or $F=Q$. In the process, $\pi'$ is the new
partition and $\pi$ the former one. Lines 7-9 consists of the
initialization of $\pi'$ to the partition of $\sim_0$, $\pi$ is
initially undefined. Lines 11-14 are the main loop of the algorithm
where $\pi$ is set to $\pi'$ and the new $\pi'$ is computed. Line 13
is described more precisely in the algorithm on the right: with each
state $p$ is associated a $k+1$-uple $s[p]$ such that two states
should be in the same part in $\pi'$ when they have the same
$k+1$-uple. The matches are found by sorting the states according to
their associated string.

The worst-case time complexity of Moore's algorithm is in $\O(n^2)$.
The following result is a more precise statement about the worst-case
complexity of this algorithm that will be used in the proof of the
main theorem (Theorem \ref{main-th}).  For sake of completeness we
also give a justification of this statement.

For any integer $n\geq 1$ and any $m\in\{0,\cdots,n-2\}$, we denote by
$\A_n^{(m)}$ the set of automata with $n$ states for which $m$ is the
smallest integer such that the $m$-equivalence $\sim_m$ is equal to
Myhill-Nerode equivalence. We also denote by $\M(\A)$ the number of
iterations of the main loop when Moore's algorithm
is applied to the automaton $\A$,

\begin{lemma}\label{moore-complexity}
For any automaton $\A$ of $\A_n^{(m)}$,
\begin{itemize}
\item the number of iterations $\M(\A)$ of the main loop in Moore's
  algorithm is at most equal to $m+1$ and always less than or equal to
  $n-1$.
\item the worst-case time complexity $\mathcal{W}(\A)$ of Moore's
 algorithm is in $\Theta((m+1)n)$, where the $\Theta$ is uniform for
 $m\in\{0,\cdots,n-2\}$, or equivalently there exist two positive real
 numbers $C_1$ and $C_2$ independent of $n$ and $m$ such that
 $C_1(m+1)n \leq {\mathcal W}(\A) \leq C_2(m+1)n$.
\end{itemize}
\end{lemma}

\begin{proof}
The result holds since the loop is iterated exactly $m+1$ times when
the set $F$ of final states is neither empty nor equal to
$\{1,\cdots,n\}$. Moreover from property (4) of
Proposition~\ref{nerode} the integer $m$ is less than or equal to
$n-2$. If $F$ is empty or equal to $\{1,\cdots,n\}$, then necessarily
$m=0$, and the time complexity of the determination of the size of $F$
is $\Theta(n)$.

The initialization and the construction of the quotient are both done
in  $\Theta(n)$.  The complexity of each iteration of the main loop is
in $\Theta(n)$: this can be achieved classically using a lexicographic
sort algorithm. Moreover in this case the constants $C_1$ and $C_2$ do
not depend on $m$, proving the uniformity of both the upper and lower
bounds.
\end{proof}

Note that Lemma~\ref{moore-complexity} gives a proof that the
worst-case complexity of Moore's algorithm is in $\O(n^2)$, as there
are no more than $n-1$ iterations in the process of the algorithm.

\subsection{Probabilistic model} \label{sec:model}
The choice of the distribution is crucial for average case analysis of
algorithms. Here we are considering an algorithm that builds the
minimal automaton of the language recognized by a given accessible
deterministic and complete one. We focus our study on the average
complexity of this algorithm for the uniform distribution over
accessible deterministic and complete automata with $n$ states, and as
$n$ tends toward infinity.  Note that for the uniform distribution
over automata with $n$ states, the probability for a given set to be
the set of final states is equal to $1/2^n$. Therefore the probability
that all states are final (or non-final) is exponentially unlikely.
Some extensions of the main result to other distributions are given in
Section~\ref{sec:discussion}.

The general framework of the average case analysis of algorithms
\cite{FS04} is based on the enumeration properties of studied objects,
most often given by generating functions. For accessible and
deterministic automata, this first step is already not easy. Although
the asymptotic of the number of such automata is known, it can not be
easily handled: a result from Korshunov~\cite{kor78}, rewritten in
terms of Stirling numbers of the second kind in~\cite{bn07} and
generalized to possibly incomplete automata in~\cite{bdn08}, is that
the number of accessible and deterministic automata with $n$ states is
asymptotically equal to $\alpha\beta^n n^{(|A|-1)n}$ where $\alpha$
and $\beta$ are constants depending on the cardinality $|A|$ of the
alphabet, and $\alpha$ depends on whether we are considering complete
automata or possibly incomplete automata.

Here some good properties of Myhill-Nerode equivalence allow us to
work independently and uniformly on each transition structure. In this
way the enumeration problem mentioned above can be
avoided. Nevertheless it should be necessary to enumerate some subsets
of this set of automata in order to obtain a more precise result. One
refers the readers to the discussion of Section~\ref{sec:conj} for
more details.

\section{Main result}\label{sec:main}

This section is devoted to the statement and the proof of the main theorem.
\begin{theorem}\label{main-th}
For any fixed integer $k\geq 1$ and for the uniform distribution over the
accessible deterministic and complete automata of size $n$ over a
$k$-letter alphabet, the average complexity of Moore's state
minimization algorithm is  $\O(n \log n)$.
\end{theorem}
Note that this bound is independent of $k$, the size of the alphabet
considered.  Moreover, as we shall see in Section \ref{sec:unary}, it
is tight in the case of a unary alphabet.

Before proving Theorem~\ref{main-th} we introduce some definitions and
preliminary results.  Let $\T$ be a fixed transition structure with
$n$ states and $\ell$ be an integer such that $1\leq\ell< n$.  Let
$p,q,p',q'$ be four states of $\T$ such that $p\neq q$ and $p'\neq
q'$.
We define $\F_{\ell}(p,q,p',q')$ as the set of sets of final states
$F$ for which in the automaton $(\T,F)$ the states $p$ and $q$ are
$(\ell-1)$-equivalent, but not $\ell$-equivalent, because of a word of
length $\ell$ mapping $p$ to $p'$ and $q$ to $q'$ where $p'$ and $q'$
are not $0$-equivalent.  In other words $\F_{\ell}(p,q,p',q')$ is the
following set:
\begin{align*}
\F_{\ell}(p,q,p',q') = \{ F\subset \{1,\cdots,n\}\mid &\ \text{for the
    automaton }(\T,F),\ p\sim_{\tiny{\ell-1}}q, \iver{p'\in F} \neq
    \iver{q'\in F}, \\ & \ \exists u\in A^\ell,\ p\cdot u=p'\text{ and }
    q\cdot u=q' \}
\end{align*}
Note that when  $\ell$ grows, the definition of $\F_{\ell}$ is more
constrained and consequently fewer non-empty sets $\F_{\ell}$ exist.

From the previous set $\F_{\ell}(p,q,p',q')$ one can define the
undirected graph $\G_{\ell}(p,q,p',q')$, called the {\em dependency
graph}, as follows:
\begin{itemize}
\item its set of vertices is $\{1,\cdots,n\}$, the set of states of
  $\T$;
\item there is an edge $(s,t)$ between two vertices $s$ and $t$ if and only
  if for all $F\in\F_{\ell}(p,q,p',q')$, $\iver{s\in F} = \iver{t\in F}$.
\end{itemize}
The dependency graph contains some information that is a basic
ingredient of the proof: it is a convenient representation of
necessary conditions for a set of final states to be in
$\F_{\ell}(p,q,p',q')$, that is, for Moore's algorithm to require more
than $\ell$ iterations because of $p$, $q$, $p'$ and $q'$. These
necessary conditions will be used to give an upper bound on the
cardinality of $\F_{\ell}(p,q,p',q')$ in Lemma~\ref{nbr-F}.

\begin{lemma}\label{acyclic}
For any integer $\ell \in \{1, \cdots, n-1\}$ and any states
$p,q,p',q'\in\{1,\cdots,n\}$ with $p\neq q$, $p'\neq q'$ such that
$\F_{\ell}(p,q,p',q')$ is not empty, there exists an acyclic subgraph
of $\G_{\ell}(p,q,p',q')$ with $\ell$ edges.
\end{lemma}
\begin{proof}

If $\F_{\ell}(p,q,p',q')$ is not empty, let $u=u_1\cdots u_\ell$ with
$u_i\in A$ be the smallest (for the lexicographic order) word of
length $\ell$ such that $p\cdot u =p'$ and $q\cdot u = q'$. Note that
every word $u$ of length $\ell$ such that $p\cdot u =p'$ and $q\cdot u
= q'$ can be used. But a non-ambiguous choice of this word $u$
guarantees a complete description of the construction.

For every
$i\in\{0,\cdots,\ell-1\}$, let $G_{\ell,i}$ be the subgraph of
$\G_{\ell}(p,q,p',q')$ whose edges are defined as follows. An edge $(s,t)$
is in $\G_{\ell,i}$ if and only if there exists a prefix $v$ of $u$ of
length less than or equal to $i$ such that $s=p\cdot v$ and $t=q\cdot
v$. In other words the edges of $\G_{\ell,i}$ are exactly the edges
$(p\cdot v,q\cdot v)$ between the  states $p\cdot v$ and $q\cdot v$
where $v$ ranges over the prefixes of $u$ of length less than or equal
to $i$. Such edges belong to $\G_{\ell}(p,q,p',q')$ since $p\sim_{\ell -
1}q$. Moreover, the graphs $\left(G_{\ell,i}\right)_{0\leq i \leq \ell-1}$
have the following properties:
\begin{enumerate}
\item For each $i\in\{0,\cdots,\ell-2\}$, $G_{\ell,i}$ is a strict
  subgraph of $G_{\ell,i+1}$. The graph $G_{\ell, i+1}$ is obtained
  from $G_{\ell,i}$ by adding an edge from $p\cdot w$ to $q\cdot w$,
  where $w$ is the prefix of $u$ of length $i+1$. This edge does not
  belong to $G_{\ell,i}$, for otherwise there would exist a strict
  prefix $z$ of $w$ such that either $p\cdot z = p\cdot w$ and $q\cdot
  z = q\cdot w$ or $p\cdot z = q\cdot w$ and $q\cdot z = p\cdot w$. In
  this case, let $w'$ be the word such that $u =ww'$, then either
  $p\cdot zw' = p'$ and $q\cdot zw' = q'$ or $p\cdot zw' = q'$ and
  $q\cdot zw' = p'$. Therefore there would exist a word of length less
  than $\ell$, $zw'$, such that, for $F \in \F_{\ell}(p,q,p',q')$,
  $\iver{p \cdot zw' \in F} \neq \iver{q'\cdot zw' \in F}$ which is
  not possible since $p\sim_{\tiny{\ell-1}}q$ and
  $\F_{\ell}(p,q,p',q')$ is not empty.  Hence this edge is a new one.
\item For each $i\in\{0,\cdots,\ell-1\}$, $G_{\ell,i}$ contains $i+1$ edges.
It is a consequence of property (1), since $G_{\ell,0}$ has only one edge
  between $p$ and $q$.
\item For each $i\in\{0,\cdots,\ell-1\}$, $G_{\ell,i}$ contains no
  loop. Indeed $p\cdot v \neq q\cdot v$ for any prefix of $u$ since
  $p\not\sim q$ for any automaton $(\T,F)$ with $F\in\F_{\ell}(p,q,p',q')$,
  which is not empty.
\item For each $i\in\{0,\cdots,\ell-1\}$, if there exists a path in
  $G_{\ell,i}$ from $s$ to $t$, then $s\sim_{\ell-1-i}t$ in every
  automata $(\T,F)$ with $F\in\F_{\ell}(p,q,p',q')$. This property can
  be proved by  induction.
\end{enumerate}

We claim that every $G_{\ell,i}$ is acyclic. Assume that it is not
 true, and let $j\geq 1$ be the smallest integer such that
 $G_{\ell,j}$ contains a cycle. By property (1), $G_{\ell,j}$ is
 obtained from $G_{\ell,j-1}$ by adding an edge between $p\cdot w$ and
 $q\cdot w$ where $w$ is the prefix of length $j$ of $u$. As
 $G_{\ell,j-1}$ is acyclic, this edge forms a cycle in
 $G_{\ell,j}$. Hence in $G_{\ell,j-1}$ there already exists a path
 between $p\cdot w$ and $q\cdot w$. Therefore by property (4) $p\cdot
 w \sim_{\ell-j} q\cdot w$ in any automaton $(\T,F)$ with
 $F\in\F_{\ell}(p,q,p',q')$.  Let $w'$ be the word such that
 $u=ww'$. The length of $w'$ is $\ell-j$, hence $p\cdot u$ and $q\cdot
 u$ are both in $F$ or both not in $F$, which is not possible since
$F\in\F_{\ell}(p,q,p',q')$.

Thus $G_{\ell,\ell-1}$ is an acyclic subgraph of $\G_{\ell}(p,q,p',q')$
with $\ell$ edges according to property (2), which concludes the proof.
\end{proof}

\begin{figure}
\begin{tabular}{cc}
\includegraphics[width=7cm]{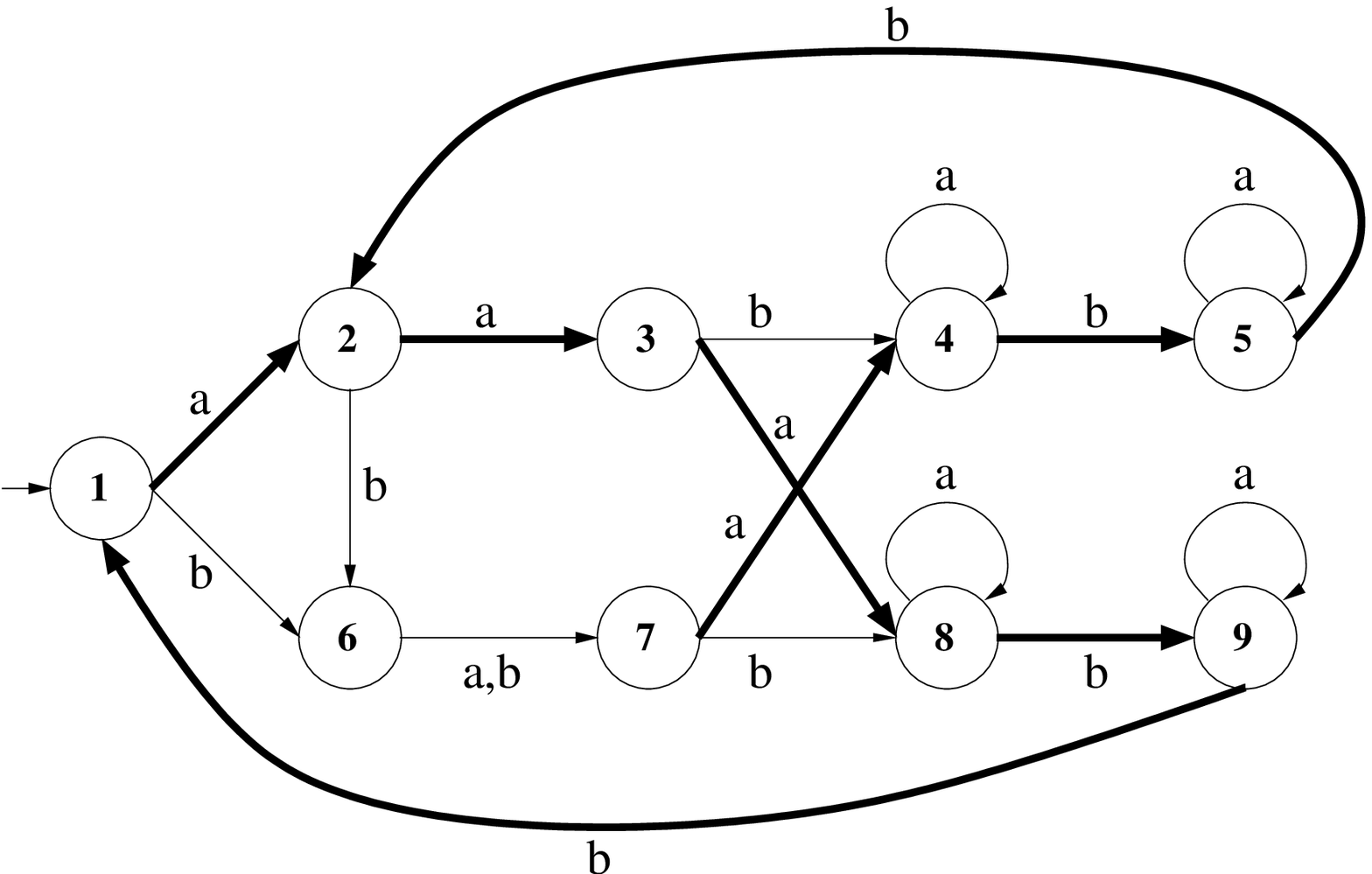} & 
\includegraphics[width=7cm]{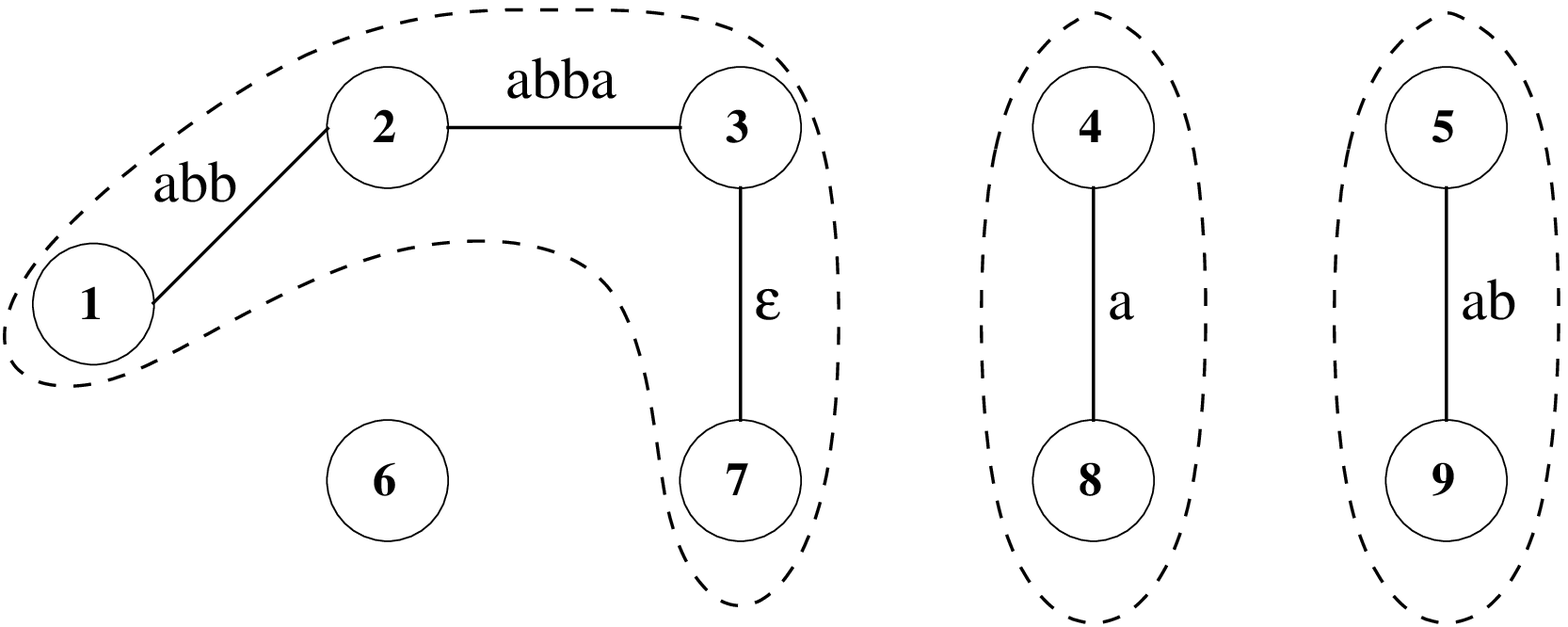}\\
\textbf{(a)} & \textbf{(b)} \\ 
\end{tabular}
\caption{Illustration of the proof of Lemma~\ref{acyclic} for $n=9$, $\ell=5$,
$p=3$, $q=7$, $p'=3$ and $q'=8$ on a given  transition structure.
\textbf{(a)} $u=abbaa$ is the smallest word of
  length $5$, for the lexicographic order, such that $3\cdot u = 3$
  and $7\cdot u = 8$. The set $\F_5(3,7,3,8)$ is not empty, as
  it contains $\{4,8\}$. The bold transitions are the ones followed
  when reading $u$ from $p$ and from $q$.
\textbf{(b)} The construction of an acyclic subgraph of
  $\G_5(3,7,3,8)$ with $5$ edges. To each strict prefix $v$ of $u=abbaa$ is
  associated an edge between $3\cdot v$ and $7\cdot v$. It encodes
  some necessary conditions for a set of final states $F$ to be in 
  $\F_5(3,7,3,8)$, as two states in the same connected component
  must be either both final or both not final. 
  }
\end{figure}

\begin{lemma}\label{nbr-F}
Given a transition structure $\T$ of size $n\geq 1$ and an integer $\ell$
with $1\leq\ell< n$, for all  states $p,q,p',q'$ of $\T$ with
$p\neq q$ and $p'\neq q'$  the following result holds:
$$|\F_{\ell}(p,q,p',q')|\leq 2^{n-\ell}.$$
\end{lemma}

\begin{proof}
If $\F_{\ell}(p,q,p',q')$ is empty, the result holds. Otherwise, from
Lemma~\ref{acyclic}, there exists an acyclic subgraph $G$ of
$\G_{\ell}(p,q,p',q')$ with $\ell$ edges. Let $m$ be the number of
connected components of $G$ that are not reduced to a single
vertex. The states in such a component are either all final or all
non-final. Therefore there are at most $m$ choices to make to
determine whether the states in those components are final or
non-final. As the graph $G$ is acyclic, there are exactly $m+\ell$
vertices that are not isolated in $G$. Hence there are at most
$2^{m}2^{n-(m+\ell)}=2^{n-\ell}$ elements in $\F_{\ell}(p,q,p',q')$:
$2^m$ corresponds to the possible choices for the connected components
and $2^{n-(m+\ell)}$ to the choices for the isolated vertices.
\end{proof}

\begin{proposition}\label{main-prop}
Let $k\geq 1$. There exists a positive real constant $C$ such that for
any positive integer $n$ and any deterministic and complete transition
structure $\T$ of size $n$ over a $k$-letter alphabet, for the
uniform distribution over the sets $F$ of final states, the average number of
iterations of the main loop of Moore's algorithm applied to $(\T,F)$ is upper 
bounded by $C \log n$.
\end{proposition}

\begin{proof}
Let $\T$ be a deterministic and complete transition structure of size
$n$ over a $k$-letter alphabet. Denote by $\F^{\geq \ell}$ the set of
sets $F$ of final states such that the execution of Moore's algorithm
on $(\T, F)$ requires more than $\ell$ iterations or equivalently such
that $(\T, F) \in \A_n^{(m)}$ with $m \geq \ell$ (see
Section~\ref{sec-moore} for  notation).

A necessary condition for $F$ to be in $\F^{\geq \ell}$ is that there exist
two states $p$ and $q$ with $p\neq q$ and such that $p\sim_{\ell-1} q$
and $p\not\sim_{\ell} q$. Therefore  there exists a word $u$ of length
$\ell$ such that $\cg p\cdot u \cd \neq\cg q\cdot u \cd$. Hence
$F\in\F_{\ell}(p,q,p\cdot u,q\cdot u)$ and
$$ \F^{\geq \ell} = \bigcup_{\substack{p,q,p',q'\in\{1,\cdots,n\}\\
p\neq q,\ p'\neq q'}} \F_{\ell} (p,q,p',q').$$ In this union the sets
$\F_{\ell} (p,q,p',q')$ are not disjoint, but this characterization of
$\F^{\geq\ell}$ is  precise enough  to obtain a useful upper bound of the
cardinality of $\F^{\geq\ell}$.  From the description of $\F^{\geq\ell}$ we
get 
$$|\F^{\geq \ell}|  \leq \sum_{\substack{p,q,p',q'\in\{1,\cdots,n\}\\ p\neq
q,\ p'\neq q'}} |\F (p,q,p',q')|, $$
and using Lemma~\ref{nbr-F} and estimating the number of choices of
the four points $p,q,p',q'$, we have    
\begin{equation}\label{eq:fgl}
 |\F^{\geq \ell}|  \leq n(n-1)\,n(n-1) 2^{n-\ell}\leq n^4
2^{n-\ell}. 
\end{equation}

For a fixed integer $\ell$ and for the uniform distribution over the
sets $F$ of final states, the average number of iterations of the main
loop of Moore's algorithm is
$$\frac1{2^n} \sum_{F\subset\{1,\dots,n\}} \M(\T,F) = \frac1{2^n}
\sum_{F\in\F^{<\ell}} \M(\T,F) + \frac1{2^n} \sum_{F\in\F^{\geq \ell}}
\M(\T,F),$$ 
where $\F^{<\ell}$ is the complement of $\F^{\geq \ell}$ in the set of
all subsets of states.  
Moreover by Lemma~\ref{moore-complexity},   for any $F\in\F^{<\ell}$,
$\M(\T,F)\leq \ell$. Therefore, since $|\F^{<\ell}|\leq 2^n$
$$\frac1{2^n} \sum_{F\in\F^{\leq\ell}} \M(\T,F)\leq \ell$$ Using
Lemma~\ref{moore-complexity} again to give an upper bound for
$\M(\T,F)$ when $F\in\F^{\geq \ell}$ and the estimate of $|\F^{\geq
  \ell}|$ given by Equation~\ref{eq:fgl} we have
$$ \frac1{2^n} \sum_{F\subset\F^{\geq \ell}} \M(\T,F) \leq 
n^5 2^{-\ell}.$$

Finally, choosing $\ell = \lceil 5\log_2 n\rceil$, we obtain that
there exists positive real $C$ such that $$ \frac1{2^n} \sum_{F
\subset \{1,\dots,n\}} \M(\T,F) \leq \lceil 5\log_2 n\rceil + n^5
2^{-\lceil 5\log_2 n\rceil} \leq C\log n,$$ concluding the proof.
\end{proof}

Now we  prove Theorem~\ref{main-th}:
\begin{mainproof}
Let $\T_n$ denote the set of deterministic, accessible and complete
transition structures with $n$ states.  For a transition
structure $\T\in\T_n$, there are exactly $2^n$ distinct automata
$(\T,F)$. 

Recall that the set $\A_n$ of deterministic, accessible and complete
automata with $n$ states is in bijection with the pairs $(\T,F)$
consisting of a deterministic, accessible and complete transition
structure $\T \in \T_n$ with $n$ states and a subset $F \subset
\{1,\cdots,n\}$ of final states.  Therefore, for the uniform
distribution over the set $\A_n$, the average number of iterations of
the main loop when Moore's algorithm is applied to an element of
$\A_n$ is
$$ \frac1{|\A_n|}\sum_{\A\in\A_n}\M(\A) =
\frac1{2^n|\T_n|}\sum_{\T\in\T_n}\sum_{F\subset\{1,\cdots,n\}} \M(\T,F)
$$
Using Proposition~\ref{main-prop} we get
$$ \frac1{|\A_n|}\sum_{\A\in\A_n}\M(\A) \leq
\frac1{|\T_n|}\sum_{\T\in\T_n}C\log n \leq C\log n.
$$
Hence the average number of iterations is bounded by $C\log n$, and by
Lemma~\ref{moore-complexity}, the average complexity of Moore's algorithm
is upper bounded by $C_1Cn\log n$, concluding the proof.
\end{mainproof}

\section{Tight bound for unary automata}\label{sec:unary}
In this section we prove that the bound $\O(n\log n)$ is optimal for
the uniform distribution on unary automata with $n$ states, that is,
automata on a one-letter alphabet.

We shall use the following result on words, whose proof is given in
detail in~\cite[p.~285]{FS04}. For a word $u$ on the binary alphabet
$\{0,1\}$, the {\em longest run of $1$} is the length of the longest
consecutive block of $1$'s in $u$.
\begin{proposition}\label{longest-run}
  For any real number $h$ and for the uniform distribution on binary
  words of length $n$, the probability that the longest run of $1$ is
  smaller than $\lfloor \log_2 n + h\rfloor$ is equal to
$$e^{-\alpha(n) 2^{-h-1}} + \O\left(\frac{\log n}{\sqrt{n}}\right),$$
where the $\O$ is uniform on $h$, and $\alpha(n) = 2^{\log n-\lfloor
  \log n \rfloor}$. 
\end{proposition}
\begin{corollary}\label{cor-longest-run}
For the uniform distribution on binary words of length $n$, the
probability that the longest run of $1$ is smaller than $\lfloor
\frac12\log_2 n \rfloor$ is smaller than $e^{-\sqrt{n}/2}$.
\end{corollary}
\begin{proof}
Set $h = -\frac{1}{2}\log_2n$ in Proposition~\ref{longest-run}
and use that for any integer $n$, $\alpha(n)\geq 1$.
\end{proof}

The shape of an accessible deterministic and complete automaton with
$n$ states on a one-letter alphabet $A=\{a\}$ is very
specific. If we label the states using the depth-first order, then for
all $q\in\{1,\cdots,n-1\}$  $q\cdot a = q+1$. The
state $n\cdot a$ entirely determines the transition structure of the
automaton. Hence there are $n2^n$ distinct unary automata with $n$
states. 
We shall also use the following result from~\cite{nic99}:
\begin{proposition}\label{minimal-unary}
For the uniform distribution on unary automata with $n$ states, the
probability that an automaton is   minimal is asymptotically equal 
to $\frac12$.
\end{proposition}

We can now prove the optimality of the $\O(n\log n)$ bound for unary
automata:
\begin{theorem}\label{optimal-unary}
For the uniform distribution on unary automata with $n$ states, the
average time complexity of Moore's state minimization algorithm is 
$\Theta(n\log n)$.
\end{theorem}
\begin{proof}
From Theorem~\ref{main-th} this time complexity is  $\O(n\log n)$.
It remains to study the lower bound of the average time complexity of
Moore's algorithm.

For any binary word $u$ of size $n$, we denote by $F(u)$ the subset of
$\{1,\cdots, n\}$ such that $i\in F(u)$ if and only if the $i$-th
letter of $u$ is $1$. The map $F$ is clearly a bijection between the binary
words of length $n$ and the subsets of $\{1,\cdots, n\}$.  Therefore a unary
automaton with $n$ states is  completely defined by a word $u$ of
length $n$, encoding the set of final states, and an integer
$m\in\{1,\cdots,n\}$ corresponding to $n\cdot a$; we denote such an
automaton by the pair $(u,m)\in \{0,1\}^n\times \{1,\cdots,n\}$. Let
$\ell$ be the integer defined by $\ell=\lfloor \frac12\log_2 n
\rfloor$. Let $M_n$ be the set of minimal unary automata with $n$
states, and $S_n$ be the subset of $M_n$ defined by $$S_n = \{(u,m)\in
M_n\mid \text{the longest run of 1 in }u\text{ is smaller than }\ell
\} $$ As the number of element in $S_n$ is smaller than the number of
automata $(u,m)$ whose longest run of 1 in $u$ is smaller than $\ell$,
from Corollary~\ref{cor-longest-run}, we have $|S_n| = o(n2^n)$. Let
$(u,m)$ be a minimal automaton in $M_n\setminus S_n$.  The word $u$
has a longest run of 1 greater or equal to $\ell$.  Let
$p\in\{1,\cdots,n\}$ be the index of the beginning of such a longest
run in $u$. The states $p$ and $p+1$ requires $\ell$ iterations in Moore's
algorithm to be separated, as $p\cdot a^i$ and $(p+1)\cdot a^i$ are
both final for every $i\in\{0,\cdots,\ell-2\}$. They must be separated
by the algorithm at some point since $(u,m)$ is minimal. Hence
$\M((u,m))\geq \ell$ for any $(u,m)\in M_n\setminus S_n$. Therefore
\begin{align*}
\frac1{n2^n}\sum_{(u,m)\in \{0,1\}^n\times\{1,\cdots,n\}}\M((u,m)) &
\geq \frac1{n2^n}\sum_{(u,m)\in M_n\setminus S_n}\M((u,m)) \\ & \geq
\frac1{n2^n}|M_n\setminus S_n|\ell \geq \frac1{n2^n}|M_n|\ell -
\frac1{n2^n}|S_n|\ell \\ & \geq \frac1{2}\ell - o(\ell)
\end{align*}
The last inequality is obtained using Proposition~\ref{minimal-unary},
concluding the proof since by hypothesis $\ell = \lfloor \frac12\log_2
n\rfloor$.
\end{proof}

\section{Extensions}\label{sec:discussion}
In this section we briefly present two  extensions of
Proposition~\ref{main-prop} and Theorem~\ref{main-th}. 

\subsection{Bernoulli distributions for the sets of final states}
Let $p$ be a fixed real number with $0<p<1$. Let $\T$ be a transition
structure with $n$ states.  Consider the distribution on the sets of
final states for $\T$ defined such that each state as a probability
$p$ of being final. The probability for a given subset $F$ of
$\{1,\cdots,n\}$ to be the set of final states is
$\mathbb{P}(F)=p^{|F|}(1-p)^{n-|F|}$.

A statement analogous to Proposition~\ref{main-prop} still holds in
this case. The proof is similar although a bit more technical, as
Proposition~\ref{main-prop} corresponds to the special case where
$p=\frac12$. Hence, for this distribution of sets of final states, the
average complexity of Moore's algorithm is also $\O(n\log n)$.

\subsection{Possibly incomplete automata}
Now consider the uniform distribution on possibly incomplete
deterministic automata with $n$ states and assume that the first step
of Moore's algorithm applied to an incomplete automaton consists in
the completion of the automaton making use of a sink state.  In this
case Proposition~\ref{main-prop} still holds.  Indeed,
Lemma~\ref{nbr-F} is still correct, even if the sets of final states
$F$ are the sets that do not contain the sink state. As a consequence,
if a transition structure $\T$ is incomplete, the average complexity
of Moore's algorithm for the uniform choice of set of final states of
the completed transition structure, such that the sink state is not
final, is in $\O((n+1)\log(n+1)) = \O(n\log n)$.

\section{Open problem}\label{sec:conj}
We conjecture that for the uniform distribution on complete,
accessible and deterministic automata with $n$ states over a
$k$-letter alphabet, with $k \geq 2$, the average
time complexity of Moore's algorithm is in $\O(n\log\log n)$.

This conjecture comes from the following observations. First,
Figure~\ref{fig:iterations} seems to show a sub-logarithmic asymptotic
number of iterations in Moore's algorithm. Second, if the automaton
with $n$ states is minimal, at least $\Omega(\log
\log n)$ iterations are required to isolate every state: $\log n$
words are needed, and this can be achieved in the best case using all
the words of length less than or equal to $\log \log n$. Moreover,
in~\cite{bn07} we conjectured that a constant part of deterministic
automata are minimal; if it is true, this would suggest that
$\Omega(\log \log n)$ is a lower bound for the average complexity of
Moore's algorithm. The conjecture above is that this lower bound is
tight.

\end{document}